\documentclass[11pt,a4paper,onehalfspacing]{article}

\usepackage{a4wide}
\usepackage{amsmath}
\usepackage{amssymb}
\usepackage{graphicx}
\usepackage{subfig}
\usepackage{dcolumn}
\usepackage{bbold}
\usepackage{bm}
\usepackage{centernot}
\usepackage{xcolor}
\usepackage{hyperref}
\usepackage{amsfonts}
\usepackage{caption}
\usepackage{hyperref}
\usepackage{authblk}

\begin{document}

\title{Entropic uncertainty relations for successive generalized measurements}
\author{Kyunghyun Baek  and Wonmin Son }
\affil{Department of Physics, Sogang University, Mapo-gu, Shinsu-dong, Seoul 121-742, Korea}
\date{}
\maketitle

\begin{abstract}
We derive entropic uncertainty relations for successive generalized measurements by using general descriptions of quantum measurement within two {distinctive operational} scenarios. In the first scenario, by merging {two successive measurements} into one we consider successive measurement scheme as a method to perform an overall {composite} measurement. In the second scenario, on the other hand, we consider it as a method to measure a pair of jointly measurable observables by marginalizing over the distribution obtained in this scheme. 
In the course of this work, we identify that limits on one's ability to measure with low uncertainty via this scheme come from intrinsic unsharpness of observables obtained in each scenario. In particular, for the L\"{u}ders instrument, disturbance caused by the first measurement to the second one gives rise to the unsharpness at least as much as incompatibility of the observables composing successive measurement.
\end{abstract}




\section{Introduction}
\label{Intro}

Ever since Heisenberg proposed uncertainty principle under consideration of $\gamma$-ray microscope in \cite{Heisenberg1927}, {the} uncertainty principle has become one of the most central concepts in quantum physics. {Till now, there have been concatenated debates to find uncertainty relations quantitatively well-formulated to reflect underlying meanings of the uncertainty principle {\cite{Busch2007}}.} {Among the uncertainty relations, one of the most widely known forms of uncertainty relations may be Robertsons's relation formulated in terms of statistical variances {\cite{Robertson1929}}. This relation was discovered by generalizing Kennard's relation {\cite{Kennard1927}} for a pair of arbitrary observables,} which indicates limitations on one's ability to prepare system being well-localized in position and momentum spaces simultaneously, so-called {\it preparation relation}. However, underlying meaning of it is not equivalent to Heisenberg's first insight that {there should be a trade-off between imprecision of an instrument measuring a particle's position and disturbance of its momentum}, which is so-called {\it error-disturbance relation}. 
From the Heisenberg's perspective, various forms of error-disturbance relation {were} derived based on state-dependent and state-independent quantifications of error and disturbance in \cite{Ozawa2003,Branciard2013} and \cite{Busch2013,Busch2014}, respectively. 
Here, the point to note in the course of {the quantifications} is that successive measurement scheme has played major roles {in clarifying} meaning of error and disturbance, and with increasing experimental ability to control quantum systems these relations {were} proved \cite{Ozawa2012,Steinberg2012} by applying this scheme. Nevertheless, uncertainty relations for successive measurements have {received} less attentions {than they deserve}, as discussed in \cite{Srinivas2002,Distler2013}.

In the field of quantum information theory, uncertainty relations have been formulated in terms of information-{theoretic} quantities such as entropy, since they are well-defined as a measure of uncertainty in the sense that they are invariant under relabeling of outcomes and concave functions (refer to \cite{Uffink1990} for further discussions). 
This information-{theoretic} approach has been conducted in both preparation and error-disturbance relations. 
From the point of preparation relation, entropic uncertainty relations {were} suggested in \cite{Deutsch1983} and then improved by Massen-Uffink \cite{Uffink1988}. A generalized version of it is written in the form of \cite{Krishna2002}
\begin{align}
H_\rho(A)+H_\rho(B)\geq c,
\end{align}
where we measure observables $A$ and $B$ described by positive-operator-valued measures (POVM) $\{\hat A_i\}$ and $\{ \hat B_j\}$ on a quantum system $\hat\rho$, respectively. In the relation, the Shannon entropy is denoted by $H(A)=-\sum_{i=1}^{n_A} p_A(i)\log p_A(i)$, where $p_A(i)=\text{Tr}[\hat A_i \hat \rho]$ is the probability to obtain $i$-th outcome of a measurement of $A$. {The} lower bound representing incompatibility between $A$ and $B$ is given by
\begin{align}\label{Incompatibility}
c=-\log\max_{i,j}\| \sqrt{\hat A_i}\sqrt{\hat B_j}\|^2,
\end{align}
where the operator norm is denoted by $\| \hat C \|$ meaning the maximal singular value of $\hat C$. Here and in the following, we will take logarithm in base 2 according to the information-{theoretic} convention.
From the point of error-disturbance relation, 
there have been recent works to formulate the relation in terms of {entropy} in order to obtain operationally meaningful formulation, based on state-dependent \cite{Coles2015-2} and state-independent quantifications
\cite{Buscemi2014}. ({see {\cite{Wehner2010}},{\cite{Coles2015}} for more details.})

{Inspired by the Heisenberg's first insight, entropic uncertainty relations for successive projective measurements were considered in an information-theoretic approach {\cite{Srinivas2002}},{\cite{Baek2014}}.  Subsequently, this approach was developed based on R\'enyi's entropies {\cite{Zhang2015}} and Tsallis' entropies {\cite{Rastegin2015}} for a pair of qubit observables. However,  the concept of generalized measurements have not been considered in successive measurement scenario. Therefore, the main purpose of the present work is to generalize the entropic uncertainty relations for the case of POVMs.
More specifically, we will focus on deriving entropic uncertainty relations for successive generalized measurements with respect to two scenarios. In the first scenario, we consider a statistical distribution of probabilities $p_{AB}(i,j)$ to obtain sequentially measurement outcomes $i$ and $j$ of the first and the second measurements $A$, $B$, respectively. In the second scenario, we analyze the marginal distributions $p_A(i)$ and $p_{B'}(j)$ associated with jointly measurable observables $A$ and $B'$, respectively.
 In particular, it was argued that the second scenario can be considered as a general method to measure any pair of jointly measurable quantum observables {\cite{Heinosaari2010}}, and further it has special usefulness due to so-called universality of successive measurement {\cite{Heinosaari2015}}. In this regard, the range of its applications becomes broader (see the references in {\cite{Heinosaari2015}}).
Additionally, in both scenarios, the effect of unsharpness of observables on entropic uncertainty relations will be discussed  by using the quantification of unsharpness previously defined in {\cite{Baek2016}}.}

This paper is organized in the following way. In Section \ref{Pre} we will introduce {a quantity defined as a measure of unsharpness} and clarify explicit mathematical expressions of measuring process. Subsequently, based on these mathematical descriptions, entropic uncertainty relations in the first scenario of successive measurement scheme will be derived in Section \ref{EUR1} with {specific} examples. In Section \ref{EUR2}, the second scenario will be considered to derive entropic uncertainty relations. Finally, we will highlight important points of the results, in Section \ref{Con}.


\section{Preliminaries}\label{Pre}

{In this section, we introduce the basic concepts necessary to generalize the entropic uncertainty relations for the case of POVMs in successive measurement scenarios.}

\subsection{Measure of unsharpness}

{Here we introduce the measure of unsharpness which {was} derived based on entropy in {\cite{Baek2016}}. 
To begin with, let us clarify notations and terminologies as follows.} 
For a finite $d$-dimensional Hilbert space $\mathcal H_{d}$, we denote the vector space of all linear operators on $\mathcal H_{d}$ by $\mathcal L(\mathcal H_d)$.
Any observable $A$ then is generally described by POVM $\{\hat A_i\}$ which is a set of positive operators $\hat A_i \in \mathcal L(\mathcal H_d)$ obeying the completeness relation, $\sum_{i=1}^{n_A} \hat A_i=\hat I$  with the number of outcomes of the measurement $n_A$. 
{In a particular case that all POVM elements are given as projections, $A$ is a projection-valued measure (PVM). In this case, the observable $A$ is commonly considered as the description of a measurement with perfect accuracy, and thus is called a {{\it sharp}} observable. On the other hand, if $A$ is not a PVM, it is called an $\text{\it unsharp}$ observable. }

{To clarify the distinction between the concepts of sharp and unsharp observables, we consider $\hat A_i$ in the form of spectral decomposition}
\begin{align}
\hat A_i=\sum_{k=1}^d a_i^k |a_i^k\rangle\langle a_i^k|,
\end{align}
where $0\leq a_i^k\leq 1$ is {an} eigenvalue corresponding to {an} eigenvector $|a_i^k\rangle$. {By means of this expression, one can find that the condition for sharp observables is equivalent to the statement} that all eigenvalues of $\hat A_i$ are given by either 0 or 1, i.e. $\forall a_i^k\in\{0,1\}$, as discussed in \cite{Busch2009}. {Otherwise, we can say that it is unsharp. This statement gives us the idea that the unsharpness measure should be defined as a function of $a_i^k$ vanishing only when $a_i^k$ is 0 or 1. Reflecting this idea, the function can be selected in the form of $h(a_i^k)=-a_i^k\log a_i^k$, and by averaging it over all POVM elements the measure of unsharpness is defined as {\cite{Baek2016}}} 
\begin{align}
D_{ \rho}(A)=\sum_{i=1}^{n_A} \sum_{k=1}^d \langle a_i^k | \hat \rho |a_i^k\rangle h(a_i^k),
\end{align}
{which is so-called $\text{{\it device uncertainty}}$, where a measurement of $A$ is performed on a quantum system $\hat \rho$. 
As a measure of unsharpness, this quantity possesses essential properties such that $D_\rho(A)=0$ for all states if and only if $A$ is a PVM, and $D_\rho(A)=H_\rho(A)$ for all states if and only if $\hat A_i=\lambda_i\hat I$ for all $i$ with $0\leq\lambda_i\leq1$ satisfying $\sum_{i=1}^{n_A}\lambda_i=1$. 
With additional properties, the validity of this quantity for the unsharpness measure has been verified on the lines of the previous work {\cite{Massar2007}} in which the unsharpness is characterized based on statistical variance.  
In particular, an important point is that the device uncertainty gives us a nontrivial lower bound of entropy by itself such that }
\begin{align}\label{Device}
H_\rho(A)\geq D_\rho(A)\geq \min_\rho D_\rho(A)\geq -\log \max_i\|\hat A_i\|,
\end{align}
{due to the concavity of entropy {\cite{Cover1991}}. Moreover, the minimal device uncertainty can be obtained by diagonalizing $\sum_{i=1}^{n_A}\sum_{k=1}^d h(a_i^k)|a_i^k\rangle\langle a_i^k|$ and taking the lowest eigenvalue, which is stronger than $-\log \max_i\|\hat A_i\|$ proposed in {\cite{Krishna2002}}. In other words, this method makes it available for us to generally find stronger state-independent bounds, and thus will play key roles in deriving entropic uncertainty relations for successive measurements. The detailed properties of the device uncertainty $D_{\rho}(A)$ has been studied in {\cite{Baek2016}}.}

\subsection{General description of successive measurements}\label{GDSM}

{In the present work, by} a {\it successive measurement}, we {mean} a scheme where two measurements are performed one after the other successively. {In particular}, the second measurement is assumed to be performed immediately on an output state conditionally {transformed} according to {an outcome of} the first measurement.
Thus, in order to consider successive measurement scheme, we should clarify how input state is transformed to output states conditioned on the measurement outcome. For this purpose, however, {the concept of} POVM is not enough to fully describe the {state transformation}. For general description of successive measurements, therefore, we need the concept of an {\it instrument} \cite{Davies1976}, which is a mapping  $\mathcal I:i\rightarrow \mathcal I_i$ such that each $\mathcal I_i$ is a completely positive linear map on $\mathcal L(\mathcal H_d)$ satisfying $\sum_{i=1}^m \text{tr}[\mathcal I_i(\hat \rho)]=1$ for all states $\hat \rho$.

\begin{figure}[t]
  \centering
    \includegraphics[width=0.8\textwidth]{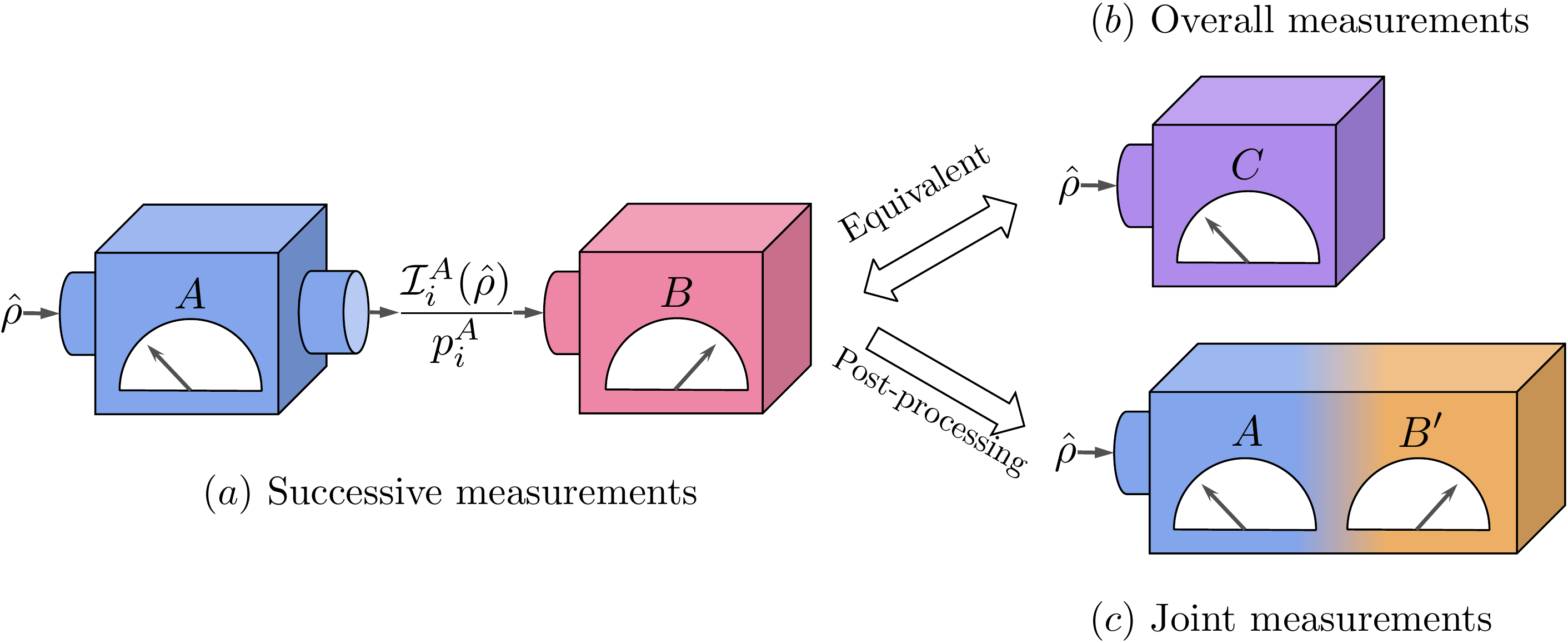}
      \caption{Relations among measurement schemes. (a) Successive measurement of observables $A$ and $B$, where the first measurement $A$ gives rise to output state $\mathcal I_i^A(\rho)/p_A(i)$ conditioned on its outcome $i$. (b) Overall measurement of $C$ obtained by performing the successive measurement of $A$ and $B$. (c) Joint measurements obtained in successive measurement scheme by considering the marginal distributions}
      \label{SM}
\end{figure}

{A general description of a quantum measurement is given by a pair of an observable $A$ and an instrument. However, a notable point is that for a given $A$ not all instruments are compatible with $A$.  For the description to be valid, an instrument should obey the condition that each $i$-th completely positive linear map $\mathcal I^A_i$ satisfies }
\begin{align}
\text{tr}[\mathcal I^A_i(\hat\rho)]=\text{tr}[\hat A_i \hat\rho]
\end{align} 
for all {states} $\hat \rho$, and in this case we say that an instrument $\mathcal I^A$ is $A$-compatible. Accordingly,  $A$-compatible instrument illustrates that a measurement outcome $i$ is obtained with the probability $p^A_i=\text{tr}[\hat A_i \hat\rho]$ for an input state $\hat \rho$, {and} a normalized output state $\mathcal I_i^A(\hat\rho)/p_i^A$ {is generated} as depicted in Figure \ref{SM}-(a). Among $A$-compatible instruments, the most common instruments occurring in applications may be the {\it L\"uders instrument}, defined by 
\begin{align}\label{Luders}
\mathcal I_{i}^{A_L}(\hat \rho)=\sqrt{\hat  A_i}\hat \rho \sqrt{\hat A_i}.
\end{align}
The L\"uders instrument is the generalized version of projective measurements for general POVM, in the sense that for sharp observables it illustrates {the same with state transformation of projective measurement.}

{Additionally, the concept of measuring process deserves consideration in order to describe successive measurement in a more specific way, of which a description is known to be consistent with the description of instrument {\cite{Ozawa1984}}. A $\text{\it measuring process}$ is defined to be a quadruple $(\mathcal{K}, |\xi\rangle, \hat U, F)$ consisting of a Hilbert space $\mathcal K$ associated with a probe system, a state vector $|\xi\rangle$ on $\mathcal K$, a unitary operator $\hat U$ on $\mathcal H\otimes \mathcal K$, and an observable $F$ on $\mathcal K$. The quadruple $(\mathcal{K}, |\xi\rangle, \hat U, F)$ is compatible with an observable $A$ if it satisfies }
\begin{align}
\text{tr}[\hat A_i\hat\rho]=\text{tr}[\hat U(\hat\rho\otimes|\xi\rangle\langle \xi|)\hat U^\dagger(\hat I\otimes \hat D_i)]
\end{align}
{for all $i$ and states. It means that the measuring process for $A$ gives rise to the same with a probability distribution obtained by performing a measurement of $A$ directly on the system. In this case, an instrument $\mathcal I^A$ is related to the measuring process $(\mathcal{K}, |\xi\rangle, \hat U, F)$ in the following manner }
\begin{align}
\mathcal I^A_i(\hat\rho)= \text{tr}_{\mathcal K}[\hat U(\hat\rho\otimes|\xi\rangle\langle \xi|)\hat U^\dagger(\hat I\otimes \hat F_i)]
\end{align}
{for all $i$ and states. In this way, each measuring process defines a unique instrument and conversely for every instrument there exists a measuring process explicitly describing the same state transformation {\cite{Ozawa1984}}. We refer to {\cite{Heinosaari2012}} for more details.}

Based on the above mathematical descriptions, there are two scenarios to investigate statistical properties of a distribution obtained in the successive measurement scheme.
Now, let us consider the first scenario in which we sequentially perform two measurements $A$ and $B$ as depicted in Figure \ref{SM}-(a), where the numbers of measurement outcomes are $n_A$ and $n_B$, respectively. Then, this scenario can be seen as a method to obtain the overall observable $C$ described by POVM $\{\hat C_{ij}\}$ obeying
\begin{align}
\text{tr}[\hat C_{ij}\hat\rho]=\text{tr}[\mathcal I^A_i(\hat\rho)\hat B_j]=p_{AB}(i,j)
\end{align}
for all $i,j$ and all states $\hat \rho$. In the Heisenberg picture, equivalently, it can be rewritten as
\begin{align}\label{C}
\hat C_{ij}=\mathcal I^{A*}_i(\hat B_j),
\end{align}
where $\mathcal I^{A*}_i$ denotes the adjoint map of $\mathcal I^A_i$. As illustrated in Figure \ref{SM}-(b), namely, the successive measurements $A$, $B$ are merged into $C$ having {$n_A n_B$} outcomes. Consequently, a task to analyze statistical {properties} in the successive measurements can be accomplished by exploring the overall observable $C$ without loss of generality. In this approach, we will investigate the entropic uncertainty relation in Section \ref{EUR1}.

On the other hand, in the second scenario we take into account the marginal distributions obtained as $\sum_{j=1}^{n_B}p_{AB}(i,j)=p_A(i)$ and $\sum_{i=1}^{n_A}p_{AB}(i,j)=p_{B'}(j)$.
Namely, the successive measurement scheme is considered as a strategy to perform a joint measurement of $A$ and $B'$ as depicted in Figure \ref{SM}-(c), where the observables $A$ and $B'$ are described by
\begin{align}\label{Joint}
\hat A_i=\sum_{j=1}^{n_B}\hat C_{ij} \;\;\;\;\; \text{ and } \;\;\;\;\; \hat B'_j=\sum_{i=1}^{n_A}\hat C_{ij}
\end{align}
for all $i$, $j$, respectively. 
It is worth noting that performing the second measurement $B$ is effectively equivalent to perform the measurement $B'$ on the initial system, since the first measurement disturbs the second one. 
However, it has been shown that, despite of this disturbance, any jointly measurable pair of quantum observables can be measured by means of successive measurement scheme \cite{Heinosaari2010}. Therefore, considering the second scenario is a general way to explicitly explore the concept of {jointly measurable observables}. Entropic uncertainty relations in this scenario will be taken into account in Section \ref{EUR2}.

\section{Generalized version of entropic uncertainty relation for successive measurements}\label{EUR1}

In the previous section, {the} mathematical methods have been presented, which are necessary to describe successive measurement in general. Based on the methods, we  derive entropic uncertainty relations within the first scenario. 
As mentioned in Section \ref{GDSM}, we can consider performing successive measurement of $A$ and $B$ as a method to {implement} the corresponding overall measurement of $C$. {This fact implies that uncertainty existing in this scenario can be equivalently characterized as entropy of $C$,}
\begin{align}
H_\rho(A,B)=H_\rho(C),
\end{align}
since $p_{AB}(i,j)=p_C(i,j)$ for all $i$, $j$. In other words, our goal to analyze uncertainty existing in the first scenario can be achieved under consideration of the overall observable $C$.
{ From this point of view, one can identify that } the reason why we cannot avoid uncertainty in this scheme originates from intrinsic unsharpness of the overall observable $C$. By quantitatively formulating this fact, as described in {Equation} \eqref{Device}, we obtain entropic form of uncertainty relation lower bounded by device uncertainty characterizing unsharpness of $C$ such that 
\begin{align}\label{EURoverall}
H_\rho(A,B)\geq D_\rho(C)\geq\min_\rho D_\rho(C)\equiv \mathcal D_1,
\end{align}
where state-independent bound $\mathcal D_1$ is obtained by minimizing device uncertainty of $C$ over all states. An important point here is that sequentially measuring incompatible observables may give rise to unavoidable unsharpness as the second measurement is disturbed by performing the first one. We can clearly observe the phenomena in the following cases.

   \subsection{Projective measurement model} \label{EUR1-pro}

In order to examine how much unsharpness emerges due to the incompatibility in the first scenario, let us consider successive projective measurements of observables $A$ and $B$ described by orthonormal bases $\{|a_i\rangle\}$ and $\{|b_j\rangle\}$ in $\mathcal H_d$, respectively. Then, according to {the state transformation} $\mathcal I_i^A(\rho)=\langle a_i|\hat \rho|a_i\rangle |a_i\rangle\langle a_i|$, this successive measurement can be considered as the overall measurement of $C$, which is described by
\begin{align}
\hat C_{ij}=|\langle a_i|b_j\rangle|^2 |a_i\rangle\langle a_i|
\end{align}
for all {$i$, $j$}. {In} this case, by calculating the minimal value of device uncertainty defined in {Equation} \eqref{Device} we obtain
\begin{align}
H_\rho(A,B)\geq \min_i \left(-\sum_{j=1}^d |\langle a_i|b_j\rangle|^2 \log |\langle a_i|b_j\rangle|^2\right),
\end{align}
where the lower bound {was proposed} in \cite{Srinivas2002}. Here, it is notable that this bound is stronger than $-\log \max_{i,j}|\langle a_i|b_j\rangle|^2$, which is widely known as a measure of incompatibility \cite{Uffink1988}. Thus, successively performing projective measurements of incompatible sharp observables induces unavoidable unsharpness which gives limits on one's ability to measure with arbitrarily low uncertainty. 

   \subsection{L\"{u}ders instrument}\label{EUR1-Luders}

As a generalized version of projective measurement model, let us assume that we implement the L\"{u}ders instrument for an unsharp observable $A$ at first and later {a} measurement of $B$ in the first scenario. {In} this case, each map $\mathcal I_i^{A_L}$ is fully determined by POVM element $\hat A_i$ as defined in {Equation} \eqref{Luders}, so that 
by applying adjoint map {of $\mathcal I^{A_L}_i$} to each POVM element of $B$ such as {Equation} \eqref{C}, we obtain {the} explicit form of the overall observable $C$ {described by} 
\begin{align}
\hat C_{ij}=\sqrt{\hat A_i}B_j\sqrt{\hat A_i}
\end{align}
for all $i$, $j$. Then, it is straightforward to formulate relations among {the concepts of} uncertainty, unsharpness and incompatibility by directly using the relations in {Equation} \eqref{Device}
\begin{align}\label{LudersOverall}
H_\rho(A,B)\geq \mathcal D_1\geq -\log \max_{i,j}\|\sqrt{\hat A_i}B_j\sqrt{\hat A_i}\|=c.
\end{align}
Here, the last inequality in {Equation} \eqref{LudersOverall} means that the minimal value of device uncertainty gives rise to {a} stronger bound than the incompatibility $c$ defined in {Equation} \eqref{Incompatibility}. {Therefore, as observed in the case of projective measurement model, one can identify that measuring incompatible observables by means of the L\"uders instrument imposes the unavoidable unsharpness.} 
In the following examples, we will analyze the relationships {in Equation {\eqref{LudersOverall}}}.

   \subsection{Examples in spin $\frac{1}{2}$ system}

\begin{figure}[t]
  \centering
    \includegraphics[width=0.32\textwidth]{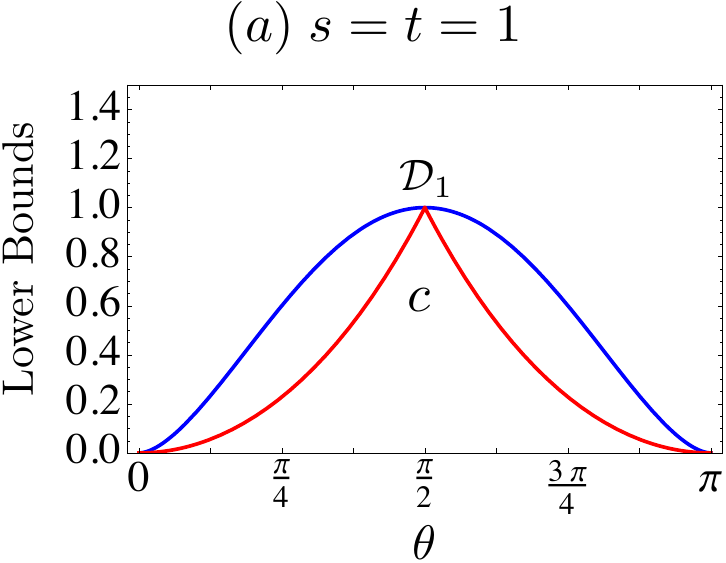} $\;$ \includegraphics[width=0.32\textwidth]{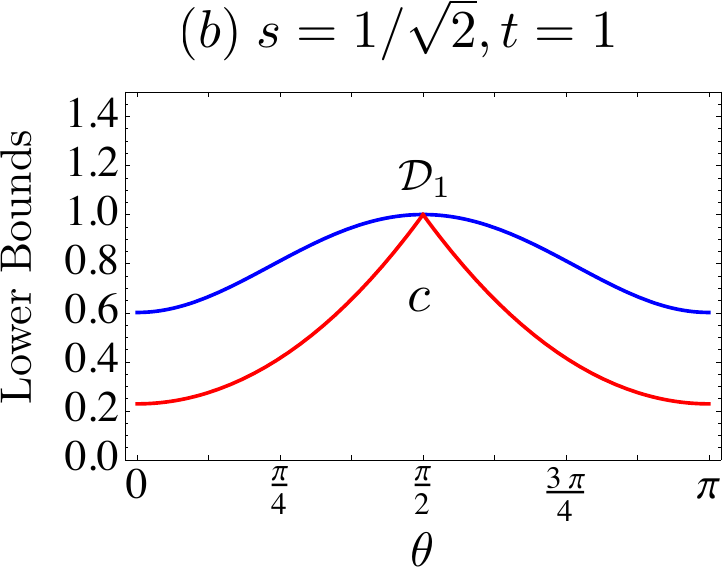} $\;$ \includegraphics[width=0.32\textwidth]{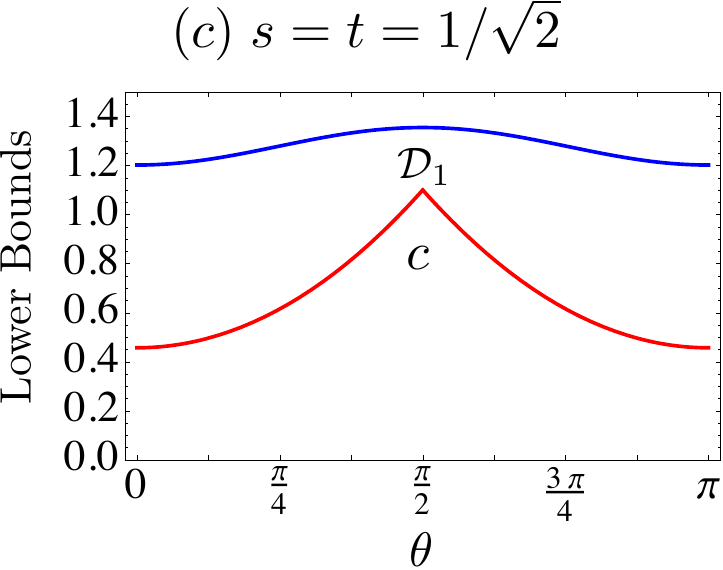}
      \caption{Graphs illustrate the lower bounds in Equation \eqref{EURoverall} for successive measurement of $Z$ and $X(\theta)$ with respect to angle $\theta$. Varying unsharpness of the observables, we present three cases. (a) At $s=t=1$, both observables are sharp. (b) At $s=1/\sqrt{2},t=1$, only the first one is unsharp. (c) At $s=t=1/\sqrt{2}$, both are unsharp.}
      \label{Graph}
\end{figure}

In order to clarify the relationship between $\mathcal D_1$ and $c$ and {verify} the validity of $\mathcal D_1$ {for} lower bound of uncertainty relation \eqref{LudersOverall},
let us consider an example of successively measuring two spin {observables} $Z$ at first and $X(\theta)$ later in $\mathcal H_2$ described by
\begin{align}
\hat Z_\pm=\frac{\hat I\pm s \hat\sigma_z}{2}\;\;\;\;\;\;\text{ and }\;\;\;\;\;\;
\hat X_\pm(\theta)=\frac{\hat I\pm t (\sin\theta\hat\sigma_x+\cos\theta\hat\sigma_z)}{2},
\end{align}
respectively, where $\hat \sigma_x$ and $\hat \sigma_z$ are the Pauli spin matrices and unsharp parameters are denoted by $0\leq s,t \leq 1$. Then the incompatibility of the observables is determined by $\theta$ which is angle between directions of measurement components. {Additionally, we assume that a $Z$-compatible instrument is induced by a measuring process $(\mathcal H_2, |\phi\rangle, \hat U_{CNOT}, \sigma_z)$, where an initial state of the probe system is $|\phi\rangle=\sqrt{(1+s)/2}|0\rangle+\sqrt{(1-s)/2}|1\rangle$ and a unitary operator $\hat U_{CNOT}$ gives rise to a CNOT gate controlled by eigenstates of $\hat\sigma_z$, $|0\rangle$ and $|1\rangle$. Namely, this measuring process leads to the L\"{u}ders instrument for $Z$.} 
{In this case, the successive} measurement scheme is equivalent to perform the overall measurement {of} $S$ defined in terms of four POVM elements
\begin{align}
\hat S_{\pm\pm}=\frac{1}{4} \Big((1 + s t \cos\theta) \hat I \pm 
    \sqrt{1 - s^2} t \sin(\theta)\; \hat \sigma_x \pm 
    (s + t \cos\theta) \hat \sigma_z\Big),\\
\hat S_{\pm\mp}=\frac{1}{4} \Big((1 - s t \cos\theta) \hat I \mp
    \sqrt{1 - s^2} t \sin(\theta)\; \hat \sigma_x \pm
    (s - t \cos\theta) \hat \sigma_z\Big).
\end{align}
Calculating the device uncertainty for the overall observable $S$, we obtain state-independent lower bounds
\begin{align}
\mathcal D_1=\sum_{\substack{\mu=\pm1 \\ \nu=\pm1}
                  }h\left(\frac{1}{4}\left(1+\mu s t \cos\theta+\nu\sqrt{s^2+t^2+\mu2st\cos\theta+s^2t^2(\cos^2\theta-1)}\right)\right).
\end{align}
We plot the lower bounds $\mathcal D_1$ and $$c=-\log\Big(\frac{1}{4}(1+st|\cos\theta|+\sqrt{s^2+t^2+2st|\cos\theta|+s^2t^2(\cos^2\theta-1)})\Big)$$ presented in {Equation} \eqref{EURoverall} versus the angle $\theta$. As a result, we can check that $\mathcal D_1$ gives rise to strictly stronger bound than $c$ except when the spin components are mutually parallel or perpendicular in Figure \ref{Graph}-(a). With increasing unsharpness of the observables, it {becomes evident} that $\mathcal D_1$ is well-formulated to reflect the effect of unsharpness as observed in Figures \ref{Graph}-(b), (c). In the examples, we analytically confirm the validity of $\mathcal D_1$ {for} a lower bound by comparing with $c$, and, {as a result, observe that} measuring incompatible observables gives rise to unavoidable unsharpness originating from the incompatibility on the assumption that we implement the L\"{u}ders instrument in the first scenario.

\section{Entropic uncertainty relations for a jointly measurable pair of observables obtained via successive measurement scheme}\label{EUR2}

In this section, we consider entropic uncertainty relations for a pair of jointly measurable observables obtained via successive measurement scheme within the second scenario. As a first step, let us clarify the concept of a joint measurement. Given observables $A$ and $B$, they are {\it jointly measurable} if and only if there exists a {\it joint observable} $M$ composed of $n_A  n_B$-elements of POVM satisfying \cite{Lahti1997}
\begin{align}
\sum_{j=1}^{n_B} \hat M_{ij}=\hat A_i\;\;\;\text{and}\;\;\;  \sum_{i=1}^{n_A} \hat M_{ij}=\hat B_j.
\end{align}
Specifically, in the second scenario, the overall observable $C$ can be seen as a joint {observable} of $A$ and $B'$ by the definitions \eqref{Joint}.
{Moreover}, as mentioned in Section \ref{GDSM}, this scheme {generally provides a method} to implement joint measurements of any jointly measurable observables \cite{Heinosaari2010}. Hence, entropic uncertainty relations obtained within this scenario is applicable to any pair of them.

Both observables $A$ and $B'$ obtained {via} the second scenario may have their own unsharpness, so that an amount of uncertainties {about} $A$ and $B'$ may not vanish due to the unsharpness of them. As formulating this fact, we obtain entropic uncertainty relations within the second scenario in the form of
\begin{align}\label{EURjoint}
H_\rho(A)+H_\rho(B')\geq D_\rho(A)+D_\rho(B')\geq \min_\rho \left[D_\rho(A)+D_\rho(B')\right]\equiv \mathcal D_2,
\end{align}
where {the sum of device uncertainties is written as} 
\begin{align}
D_\rho(A)+D_\rho(B')=\text{tr}\left[\hat \rho\left(\sum_{i=1}^{n_A}\sum_{k=1}^{d}h(a_i^k)|a_i^k\rangle\langle a_i^k| +\sum_{j=1}^{n_B}\sum_{l=1}^{d} h({b'}_j^l)|{b'}_j^l\rangle\langle {b'}_j^l|\right)\right],
\end{align}
{and thus minimizing it} over all states can be accomplished by diagonalizing {$$\big(\sum_{i=1}^{n_A}\sum_{k=1}^{d}h(a_i^k)|a_i^k\rangle\langle a_i^k| +\sum_{j=1}^{n_B}\sum_{l=1}^{d} h({b'}_j^l)|{b'}_j^l\rangle\langle {b'}_j^l|\big)$$} and taking the lowest eigenvalue. 
An important point, here, is that the second measurement $B$ may be perturbed to be $B'$ because of disturbance caused by the first measurement $A$, while $A$ is preserved. This fact implies that even when both observables $A$ and $B$ applied in this scheme are sharp, it is possible for the perturbed one $B'$ to become unsharp. In particular, this behavior is apparently observed when applying a pair of incompatible observables to this scenario, since measuring one of a pair of incompatible observables disturbs the other, according to the Heisenberg's insight. 
{From the point of view} that the incompatibility imposes {unavoidable} unsharpness on {the} observables $A$ and $B'$, we will discuss more details in the following specific measurement models.

   \subsection{Projective measurement model}\label{EUR2-Pro}

In the same way in Section \ref{EUR1-pro}, let us consider projective measurements {of} $A$ and $B$ described in $\mathcal H_d$ by orthonormal bases $\{|a_i\rangle\}$ and $\{|b_j\rangle\}$, respectively. Then, in the second scenario, the first one $A$ remains itself, while the second one $B$ is disturbed to be $B'$ described by
\begin{align}
\hat B_j'= \sum_{i=1}^d |\langle a_i|b_j\rangle|^2 |a_i\rangle\langle a_i|
\end{align}
for all $j$. {According to the L\"uders theorem {\cite{Luders1951}}, $B$ is not disturbed if and only if all elements of $A$ and $B$ commute each other.} {In} this case, thus, even though we perform two sharp observables sequentially, the second one {involves the unavoidable} unsharpness originating from  the incompatibility between them. This behavior can be quantitatively formulated in the form of
\begin{align}\label{EUR2Pro}
H_\rho(A)+H_\rho(B')\geq D_\rho(B')\geq \min_\rho D_\rho(B')=\min_i \left(-\sum_{j=1}^d |\langle a_i|b_j\rangle|^2\log |\langle a_i|b_j\rangle|^2\right),
\end{align}
where its lower bound {was proposed} in \cite{Srinivas2002}. Generalized version of it for POVMs will be taken into account in the following.

   \subsection{L\"{u}ders instruments}\label{EUR2-Luders}

As a next step, in the same manner as Section \ref{EUR1-Luders}, let us assume we sequentially implement the L\"{u}ders instrument of an observable $A$ at first and later another one of $B$, in the second scenario. Then, according to {Equation} \eqref{Luders}, it is equivalent to implement joint measurement of a pair of observables $A$ and $B'$ described by
\begin{align}\label{Luders2}
\hat A_i \;\;\;\;\;\text{ and }\;\;\;\;\; \hat B'_j = \sum_{i=1}^{n_A} \sqrt{\hat A_i} \hat B_j \sqrt{\hat A_i},
\end{align}
for all $i$, $j$, respectively. {In} this case, likewise as discussed above, performing the L\"{u}ders instrument of $A$ gives rise to disturbance to $B$, in a way for $B$ to become $B'$. Then, by applying the relations in {Equation} \eqref{Device} directly, we obtain 
\begin{align}\label{LudersJoint}
H_\rho(A)+H_\rho(B')\geq D_\rho(A)+D_\rho(B')\geq \min_\rho \left[D_\rho(A)+D_\rho(B')\right]\geq -\log \max_j\| \sum_{i=1}^{n_A} \sqrt{\hat A_i} \hat B_j \sqrt{\hat A_i}\|,
\end{align}
where {the last inequality follows from applying the third inequality in Equation {\eqref{Device}} solely to the observable $B'$ and its} bound is a new form of incompatibility larger than $-\log \max_{i,j}\|\sqrt{\hat A_i}\sqrt{\hat B_j}\|^2$, which {was} conjectured in \cite{Tomamichel2012} and proved later in \cite{Coles2014}. A distinct point from projective measurement model is that there is a possibility for $A$ to be unsharp, and loosing sharpness of $A$ may decrease disturbance to $B$ caused by $A$. Namely, {a trade-off between the unsharpness of $A$ and $B$ can be observed.} In the following examples, {this phenomenon} will be {examined}.

   \subsection{Examples in spin $\frac{1}{2}$ system}

\begin{figure}[t]
  \centering
    \includegraphics[width=0.4\textwidth]{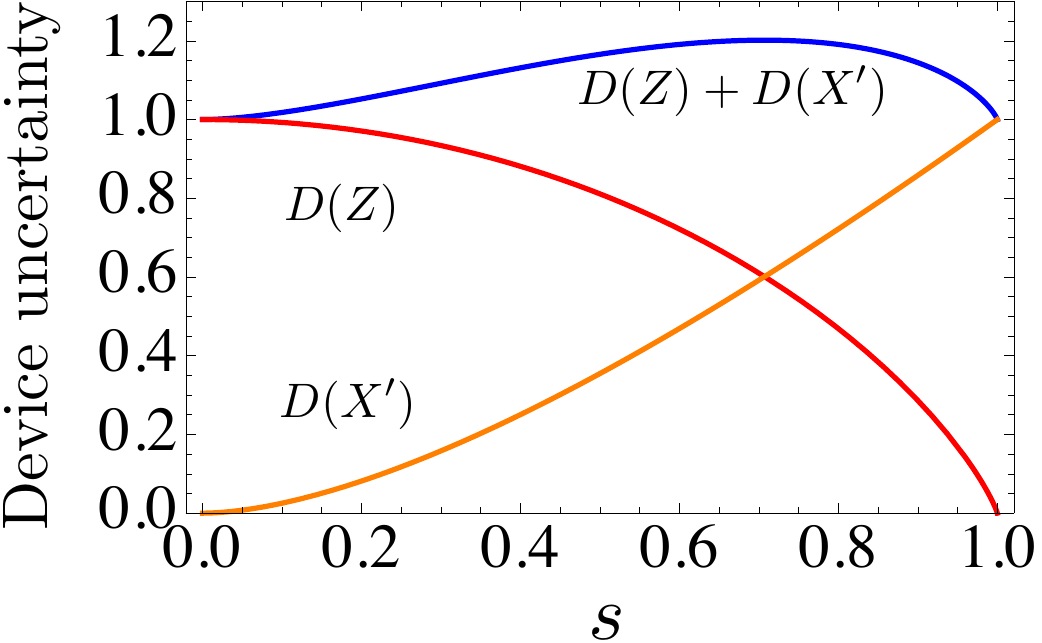}
      \caption{For a pair of jointly measurable observables $Z$ and $X'$ obtained via successive measurement scheme, we plot device uncertainties of them and their summation denotes by $D(Z)$, $D(X')$ and $D(Z)+D(X')$, respectively, versus the unsharp parameter $s$. We can observe trade-off relation between $D(Z)$ and $D(X')$ characterizing unsharpness of $Z$ and $X'$. }
      \label{Graphjoint}
\end{figure}

As an example in the second scenario, we assume to implement  the L\"{u}ders instrument of $Z$ {induced by the measuring process $(\mathcal H_2, |\phi\rangle, \hat U_{CNOT}, \sigma_z)$} and {a} measurement of $X$ {successively} in $\mathcal H_2$ described by 
\begin{align}
\hat Z_\pm=\frac{\hat I\pm s \hat\sigma_z}{2}\;\;\;\;\;\;\text{ and }\;\;\;\;\;\;
\hat X_\pm=\frac{\hat I\pm \hat\sigma_x}{2},
\end{align}
respectively. Here we restrict ourselves for $X$ to be sharp, in order to observe more clearly unsharpness {appearing due to} the first measurement. 
{In} this case, we can consider it as a method to measure a pair of jointly measurable observables $Z$ and $X'$, where the perturbed observable $X'$ is given as \cite{Carmeli2012}
\begin{align} 
\hat X'_\pm=\frac{\hat I\pm t \hat\sigma_x}{2}
\end{align}
with the unsharp parameter $t=\sqrt{1-s^2}$. Consequently, this scheme provides {an} optimized method to jointly measure incompatible observables $\sigma_z$ and $\sigma_x$, in the sense that the unsharp parameters $s$ and $t$ saturate the inequality $s^2+t^2\leq 1$, which is a necessary and sufficient condition 
for $Z$ and $X'$ to be jointly measurable \cite{Busch1986}. However, even in {the} optimized method, we can not avoid unsharpness, and {there is the trade-off between the unsharpness of $Z$ and $X'$ such} that the more sharpness of $Z$, the more unsharpness of $X'$. This behavior can be examined quantitatively by considering entropic uncertainty relations \eqref{EURjoint} given in the form of 
\begin{align} 
H_\rho(Z)+H_\rho(X')\geq D(Z)+D(X')= H_{bin}\left(\frac{1+s}{2}\right)+H_{bin}\left(\frac{1+\sqrt{1-s^2}}{2}\right),
\end{align}
where binary entropy is denoted by{$H_{bin}(q)=-h(q)-h(1-q)$}. The trade-off between {the} unsharpness of them is illustrated in Figure \ref{Graphjoint}, and the total unsharpness {characterized by} $D(Z)+D(X')$ is maximized when unsharpness equally distributed $s=t=1/\sqrt{2}$, while minimized at extreme points $s=1$ or $s=0$.

\section{Conclusion}\label{Con}

The main purpose of {present work} is to suggest entropic uncertainty relations for successive generalized measurement within two distinctive scenarios. Before deriving the relations, in Section \ref{Pre}, we have introduced device uncertainty as a measure of unsharpness and its mathematical properties in order to investigate the effect of unsharpness on entropic uncertainty relations in successive measurement scheme. Subsequently, we have explicitly explained general description of successive measurement scheme with respect to two scenarios. In the first scenario, we have considered this scheme as a method to implement an overall measurement, and, as a result, observed that unsharpness of the overall observable gives limits on one's ability to measure it with arbitrarily low uncertainty as formulated in {Equation} \eqref{EURoverall}. Assuming to perform the L\"{u}ders instrument as the first measurement in this scenario, it {is} clearly shown that this unsharpness comes from disturbance caused by the first measurement to the second one. The amount of unsharpness appears at least as much as the incompatibility of a pair of observables composing successive measurement, as formulated in {Equation} \eqref{LudersOverall}. In the second scenario, on the other hand, this scheme has been considered as a method to measure a pair of jointly measurable observables.
Consequently, we have figured out that total unsharpness in both observables is a major factor that gives rise to unavoidable uncertainty as observed in {Equation} \eqref{EURjoint}.
Also under the assumption for the first measurement to be described by the L\"{u}ders instrument, it becomes clear that the first measurement leads to unsharpness of the second one by disturbing it at least as much as incompatibility as shown in {Equation} \eqref{LudersJoint}. 
It is {notable} that this form of uncertainty relations is applicable to any pair of jointly observables, since Heinosaari {\it et al.} have proved that we can obtain any pair of jointly measurable observables via successive measurement scheme in \cite{Heinosaari2010}.

\vspace{6pt}  


\section*{Acknowledgments} 
This work was done with support of ICT R\&D program of MSIP/IITP (No.2014-044-014- 002),  the R\&D Convergence Program of NST (National Research Council of Science and Technology) of Republic of Korea (Grant No. CAP-15-08-KRISS) and National Research Foundation (NRF) grant (No.NRF- 2013R1A1A2010537).





\section*{\noindent Abbreviations}\vspace{6pt}\noindent 
The following abbreviations are used in this manuscript:\\

\noindent POVM: Positive Operator Valued Measure\\
PVM: Projector Valued Measure\\



\begin{thebibliography}{999} 

\bibitem{Heisenberg1927} Heisenberg, W. \"{U}ber den anschulichen Inhalt der quantentheoretischen Kinematik und Mechanik. {\it Z. Phys.} {\bf 1927} {\it 43}, 172-198.
\bibitem{Busch2007} Busch, P.; Heinosaari, T.; Lahti, P. Heisenberg's uncertainty principle. {\it Physics Reports} {\bf 2007} {\it 452}, 155-176.
\bibitem{Robertson1929} Robertson, H. P. The uncertainty principle {\it Phys. Rev.} {\bf 1929} {\it 34}, 163.
\bibitem{Kennard1927} Kennard, E.H. Zur Quantenmechanik einfacher Bewegungstypen. {\it Z. Phys.} {\bf 1927}, {\it 44}, 326-352 
\bibitem{Ozawa2003} Ozawa, M. Universally valid reformulation of the Heisenberg uncertainty principle on noise and disturbance in measurement, {\it Phys. Rev. A} {\bf 2003}, {\it 67}, 042105.

\bibitem{Branciard2013} Branciard, C. Error-tradeoff and error-disturbance relations for incompatible quantum measurements {\it Proc. Nat. Acad. Sci.} {\bf 2013} {\it 110}, 6742-6747.
\bibitem{Busch2013} Busch, P.; Lahti, P.; Werner, R. F. Proof of Heisenberg's error-disturbance relation, {\it Phy. Rev. Lett.} {\bf2013} {\it 111}, 160405.
\bibitem{Busch2014}  Busch, P.; Lahti, P.; Werner, R. F. Heisenberg uncertainty for qubit measurements {\it Phy. Rev. A} {\bf2014} {\it 89}, 012129.
\bibitem{Ozawa2012} Erhart, J.; Spona, S.; Sulyok, G.; Badurek, G.; Ozawa, M.; Yuji, H. Experimental demonstration of a universally valid error-disturbance uncertainty relation in spin measurements. {\it Nature Phys.} {\bf2012} {\it 8}, 185-189.
\bibitem{Steinberg2012} Rozema, L. A.; Darabi, A.; Mahler, D. H.; Hayat, A.; Soudagar, Y.; Steinberg, A. M. Violation of Heisenberg's measurement-disturbance relationship by weak measurements. {\it Phys. Rev. Lett.} {\bf 2012} {\it 109}, 100404.


\bibitem{Srinivas2002} Srinivas, M. D. Optimal entropic uncertainty relation for successive measurements in quantum information theory {\it Paramana-J. Phys.} {\bf 2002} {\it 60} 1137-1152.
\bibitem{Distler2013} Distler, J.; Paban, S. Uncertainties in successive measurements {\it Phy. Rev. A} {\bf 2013} {\it 87}, 062112.
\bibitem{Uffink1990} Uffink, J.B.M. {\it Measures of Uncertainty and the Uncertainty Principle}, Ph.D. thesis, University of Utrecht, Utrecht, Netherlands,1990.
\bibitem{Deutsch1983}Deutsch, D. Uncertainty in Quantum Measurements {\it Phy. Rev. Lett.} {\bf 1983} {\it 50}, 631.
\bibitem{Uffink1988} Maasen, H.; Uffink, J.B.M. Generalized entropic uncertainty relations {\it Phy. Rev. Lett.} {\bf 1988} {\it 60}, 1103-1106.

\bibitem{Krishna2002} Krishna M.; Parthasarathy, K.R. An entropic uncertainty principle for quantum measurements {\it Indian J. Stat. Ser. A} {\bf 2002} {\it 64}, 842
\bibitem{Coles2015-2} Coles, P.J.; Furrer, F. State-dependent approach to entropic measurement-disturbance relations. {\it Physics Letters A} {\bf 2015} {\it 379} 105-112
\bibitem{Buscemi2014} Buscemi, F.; Hall, M.J.W.; Ozawa, M.; Wilde, M.M. Noise and disturbance in quantum measurements: An information-theoretic approach. {\it Phys. Rev. Lett.} {\bf2014} {\it 112}, 050401.
\bibitem{Wehner2010} Wehner, S.; Winter, A. Entropic uncertainty relations-a survey {\it New. J. Phys.} {\bf2010} {\it12}, 025009.
\bibitem{Coles2015} Coles, P.J.; Berta, M.; Tomamichel, M.; Wehner, S. Entropic uncertainty relations and their applications. {\it Rev. Mod. Phys.} {\bf 2017} {\it 89}, 015002.
\bibitem{Baek2014} Baek, K.; Farrow, T.; Son, W. Optimized entropic uncertainty for successive projective measurements {\it Phys. Rev. A} {\bf2014} {\it89}, 032108.


\bibitem{Zhang2015} Zhang, L; Zhang, Y.; Yu, C. R\'{e}nyi entropy uncertainty relation for successive projective measurements. {\it Quantum Information Processing}, {\bf 2015} {\it14}, 2239-2253.
\bibitem{Rastegin2015} Rastegin, A.E. Uncertainty and certainty relations for successive projective measurements of a qubit in terms of Tsallis' tntropies. {\it Commun. Theor. Phys.} {\bf 2015} {\it63}, 687.
\bibitem{Heinosaari2010} Heinosaari, T.; Wolf, M. Non-disturbing quantum measurements. {\it J. Math. Phys.} {\bf2010} {\it 51}, 092201.
\bibitem{Heinosaari2015} Heinosaari, T.; Miyadera, T. Universality of sequential quantum measurements {\it Phys. Rev. A} {\bf2015} {\it 91}, 022110.
\bibitem{Baek2016} K. Baek, W. Son, Unsharpness of generalized measurement and its effects in entropic uncertainty relations. {\it Sci. Rep.} {\bf 2016}, {\it 6}, 30228

\bibitem{Busch2009} Busch, P. On the sharpness and bias of quantum effects. {\it Found. Phys.} {\bf2009} {\it 39}, 712.
\bibitem{Massar2007} Massar, S. Uncertainty relations for positive-operator-valued measures. {\it Phys. Rev. A} {\bf 2007} {\it 76}, 042114.
\bibitem{Cover1991} Cover, T.M. and Thomas, J.A. {\it Elements of Information Theory}; Wiley: New York, USA, 1991.
\bibitem{Davies1976} Davies, E.B. {\it Quantum theory of open systems}; Academic Press: London, UK, 1976.
\bibitem{Ozawa1984} Ozawa, M. Conditional expectation and repeated measurements of continuous  quantum observables. {\it J. Math. Phys.} {\bf 1984} {\it 25}, 79-87
\bibitem{Heinosaari2012} Heinosaari, T. and Ziman, M. {\it The mathematical language of quantum theory from uncertainty to entanglement}; Cambridge University Press: Cambridge, UK, 2012.
\bibitem{Lahti1997} Lahti, P.; Pulmannov\'{a}, S. Coexistent observables and effects in quantum mechanics. {\it Rep. Math. Phys.} {\bf 1997} {\it39}, 339.
\bibitem{Luders1951} L\"uders, G.  \"Uber die Zustands\"anderung durch den Messprozess. {\it Ann. Physik (6)}, {\bf 1951} 8, 322–328.
\bibitem{Tomamichel2012} Tomamichel, M. A Framework for Non-Asymptotic Quantum Information Theory, Ph.D. thesis, ETH Zurich, Zurich, Switzerland, 2012.


\bibitem{Coles2014} Coles, P.J.; Piani, M. Improved entropic uncertainty relations and information exclusion relations. {\it Phys. Rev. A}, {\bf 2014} {\it 89}, 022112.
\bibitem{Carmeli2012} Carmeli, C.; Heinosaari, T.; Toigo, A. Informationally complete joint measurements on finite quantum systems. {\it Phys. Rev. A} {\bf 2012} {\it 85}, 012109.
\bibitem{Busch1986} Busch, P. Unsharp reality and joint measurements for spin observables. {\it Phys. Rev. D}, {\bf 1986} {\it 33}, 2253.




\end{thebibliography}


\end{document}